\documentstyle[12pt,aps,epsf,preprint]{revtex}
\newcommand{\ds}{\displaystyle}
\newcommand{\dsf}{\ds\frac}
\newcommand{\Tr}{\mbox{Tr}}
\newcommand{\re}[1]{(\ref{#1})}
\newcommand{\no}{\nonumber}
\newcommand{\onehalfspace}{\renewcommand{\baselinestretch}{1.}\large\normalsize}
\onehalfspace
\draft
\title {\LARGE\bf  Nucleon deformation in  finite nuclei}
\author{\small\bf U.T. Yakhshiev$^{1,2}$\footnote{
Electronic address:~ulug@iaph.post.tps.uz},
M.M. Musakhanov$^2$,  A.M. Rakhimov$^3$,\\ Ulf-G. Mei{\ss}ner$^1$\footnote{
Electronic address:~u.meissner@fz-juelich.de}, A.Wirzba$^1$\footnote{
Electronic address:~a.wirzba@fz-juelich.de}
}
\address{$^1$Forschungszentrum J{\" u}lich, Institut f{\" u}r Kernphysik
(Theorie),  D-52425  J{\" u}lich, Germany,\\
 $^2$Theoretical Physics Department and Institute of Applied Physics\\
National University of Uzbekistan, Tashkent-174, Uzbekistan\\
$^3$Institute of Nuclear Physics, Academy of Science, Uzbekistan}

\begin{document}
\setcounter{page}{0}
\begin{titlepage}
\maketitle
\thispagestyle{empty}
\centerline{\today}

\begin{abstract}
  The deformation of a nucleon embedded in various finite nuclei is considered
  by taking into account the distortion of the chiral profile functions under
  the action of an external field representing the nuclear density.  The
  baryon charge distribution of the nucleon inside light, medium--heavy and
  heavy nuclei is discussed.  The mass of the nucleon decreases as it is
  placed deeper inside the nucleus and reaches its minimum at the center of
  the nucleus. We discuss the quantization of non-spherical solitons and its
  consequences for the mass splitting of the $\Delta$ states.  We show that
  bound nucleons acquire an intrinsic quadrupole moment due to the deformation
  effects. These effects are maximal for densities of nuclei about
  $\rho(R)\sim 0.3\div 0.35\rho(0)$.  We also point out that scale changes of
  the electromagnetic radii can not simply be described by an overall swelling
  factor.
\end{abstract}

\vskip 5mm

\vskip 3mm
%\keywords{
%Keywords:
%Skyrme model, nuclear medium,
%soliton deformation, nucleon quadrupole moment.
%}

\end{titlepage}

%%%%%%%%%%%%%%%%%%%%%%%%%%%%%%%%%% section 1
\section{Introduction}

The possible modification of nucleon properties in the nuclear medium is
currently a much discussed topic in low energy hadron physics. One way to
consider such problems is to describe the nucleon as a topological soliton and
then study the influence of baryonic matter on the properties of such
solitons, making use of the independent particle picture which has been so
successful in describing many properties of nuclei. There exist already some
works where nucleons described as Skyrme--type solitons embedded in infinite
nuclear matter have been considered~\cite{prc001,npa001}.\footnote{For an
  earlier and somewhat different approach to density effects in Skyrme--type
  models, which leads to similar results, see ref.~\cite{UGM}. Some other
  works concerned with nuclear matter aspects of the Skyrme model are
  collected in ref.~\cite{matter}.}  The results of these studies were in
qualitative agreement with experimental indications and with results of other
authors using different approaches. On the quantitative level, however, there
is a too large renormalization of the nucleons' effective mass in nuclear
matter.  The quantitative value of this renormalization is about $\sim 40\%$
for the normal nuclear matter density, $\rho_0=0.5m_\pi^{-3}$~\cite{prc001}.
It is therefore difficult to relate this modification of the nucleon
self--energy in the medium to the nucleon mass in free space because such
calculations consider only the special case when a nucleon is placed in the
center of the heavy nucleus.  The infinite nuclear matter approach allows one
only to consider properties of nucleons placed near the center of heavy nuclei
where the density is constant.  One can expect that taking into account
non-spherical effects, i.e. deformation of the skyrmion in the finite nucleus,
would improve the results also on the quantitative level.  Density changes
play an important role when the nucleon is placed at sufficiently large
distances from the center of nucleus.

The role of deformation effects (for an early work see~\cite{Jackson})
 has already been investigated in connection
with the nucleon--nucleon interaction in ref.~\cite{KM}.  The results
presented in this work demonstrated the necessity of non-spherical deformed
solutions of the field equations to obtain a reliable description of the $NN$
potential.  A similar approach to the deformation for two interacting
skyrmions has been used in the work~\cite{Rakhimov96}, where changes in the
nucleon shape are investigated by letting the nucleon deform under the strong
interactions with another nucleon.  On the other hand, rotational and
vibrational excitations of deformed skyrmions have also been considered for
baryon number equal to one ($B=1$) system~\cite{Hajduk}. In a series of
papers~\cite{Kopeliovich,Nikolaev,Otofuji} axially symmetric few--skyrmion
systems with baryon number $B$ larger than one were investigated, because
spherically symmetric configurations in the skyrme model do not give bound
skyrmionic systems~\cite{Bogomolniy} even for the solutions of the model
within the finite nuclei as considered in ref.~\cite{ujf003} (for similar or
related work, see also ref.~\cite{shape}).  In the
works~\cite{Hajduk,Nikolaev,Otofuji} the deformation is introduced by a
modification of the unit spherical vector and additionally through the
modification of the profile function~\cite{Kopeliovich} in the expression of
the ansatz for the chiral field (for more precise definitions, see below). The
results of all investigations for $B>1$ systems lead to the conclusion that
the system has to be deformed to obtain the ground state. Some other papers
that deal with deformation effects and non-spherical configurations are
collected in refs.~\cite{deformax}.

In the present work we consider properties of the deformed nucleon embedded in
light, medium--heavy and heavy nuclei. Deformation effects are introduced by
the distortion of the profile function of chiral field under the action of the
external field (which parameterizes the baryonic density within a given
nucleus). In contrast to most previous investigations, we do not consider a
constant nuclear density as appropriate for nuclear matter but rather the
distance-dependent density as given for finite nuclei.  We calculate the
modifications of the nucleon properties in finite nuclei, and their dependence
on the distance between the topological center of the skyrmion inside the
nucleus and the geometrical center of the nucleus under consideration will be
considered.

The paper is organized in the following way. After the formulation of the
problem in section~\ref{sec:form}, we discuss the classical soliton mass and
distribution of baryon charge for a skyrmion inside finite nuclei in
section~\ref{sec:mass}. Based upon the results for the mass and baryon charge
distribution of the deformed classical soliton, we present in
section~\ref{sec:quant} the quantization of axially symmetric systems with
baryon number $B=1$.  Section~\ref{sec:prop} is devoted to the quantum
properties of the solitons, in particular the effective mass of nucleons in
finite nuclei and the consequences for $\Delta$ mass splittings. We also
consider the density dependence of the nucleons electromagnetic radii and
calculate the intrinsic quadrupole moment of the deformed nucleon. Finally, in
section~\ref{sec:concl} the conclusions and an outlook for further
investigations are given.

%%%%%%%%%%%%%%%%%%%%%%%%%%%%%%%%%% section 2
\section{Formulation of the problem}
\label{sec:form}
%%%%%%%%%%%%%%%%%%%%%%%%%%%%%%%%%% subsection 2.1
\subsection{Lagrangian and ansatz}

Our starting point is the medium--modified Skyrme lagrangian~\cite{prc001}
\begin{equation}
\begin{array}{l}
{\cal L}=
\dsf{F_{\pi}^2}{16}\mbox{Tr}\left(\dsf{\partial{U}}{\partial t}\right)
\left(\dsf{\partial{U}^\dagger}{\partial t}\right)-
\dsf{F_{\pi}^2}{16}\alpha_p(\vec r)\mbox{Tr}
(\vec\nabla{U})(\vec\nabla{U^\dagger})\\
\quad \\
\qquad \qquad
+\dsf{1}{32e^2}\mbox{Tr}[L_{\mu},L^{\nu}]^2
+\dsf{F_{\pi}^2m_{\pi}^2}{16}\alpha_s(\vec r)\mbox{Tr}[U+U^{\dagger}-2]\,\,,
\label{lag}
\end{array}
\end{equation}
where $L_\mu=U^+\partial_\mu U$, $F_\pi = 186\,$MeV is the pion decay
constant, $e$ is a dimensionless parameter and $m_\pi = 138\,$MeV is the pion
mass.  The medium functionals $\alpha_s(\vec r)$ and $\alpha_p(\vec r)$ are
expressed via the functionals $\chi_s(\vec r)$ and $\chi_p(\vec r)$ of
$s$-wave and $p$-wave pion--nucleon scattering lengths and the nuclear density
$\rho(\vec r \,)$:
\begin{equation}
\alpha_p(\vec r \,)=1-\chi_p(\vec r \,)\,\,,\qquad
\alpha_s(\vec r \,)=1+\dsf{\chi_s(\vec r \,)}{m_{\pi}^2}\,\,.
\label{medfunc}
\end{equation}
Here, $\chi_p(\vec r)$ has the meaning of the pion dipole susceptibility of
the medium.  The influence of the medium enters through these functionals (for
explicit expressions, see section~\ref{sec:sus}).\footnote{Such an approach
  has also been considered in next-to-leading order chiral perturbation theory
  in which the pions are chirally coupled to matter fields, like e.g.
  nucleons, see Ref.~\cite{AW}.} Note that we employ here the same
approximations as in~\cite{prc001}. Since the fourth order term can be
obtained from the infinite mass limit of $\rho$ meson exchange, a possible
density dependence of that term could enter via in--medium modified $\rho$
meson properties as it was done e.g. in~\cite{UGM}. Similarly, one could add a
sixth order term $\sim B_\mu B^\mu$ arising from $\omega$ meson exchange or an
symmetric fourth order term due to scalar mesons. We do not consider such
terms here. Also, since the empirical information on possible changes of
vector meson properties in the medium has not yet lead to an established
picture, we treat the coupling $e$ as a density independent number. This could
be released in more sophisticated model but goes beyond the scope of the
present investigation.  For a more detailed discussion about how the specific
form of the  Lagrangian in eq.~\re{lag} arises, we refer to
ref.~\cite{prc001}.

The ansatz for the chiral field  $U$ has the hedgehog form
\begin{equation}
U(\vec r)=\exp\left[\dsf{i\vec\tau \cdot \vec r}{|\vec r\,|}F(|\vec r\,|)
\right]
\label{anzsph}
\end{equation}
for the skyrmion placed at the center of the nucleus.

When the  skyrmion is placed at a distance $\vec R$ from the center
of the spherical nucleus (see Fig.~\ref{fig1}), the chiral field $U$
can be written in the form
\begin{equation}
U(\vec r-\vec R)=\exp\left[i\vec\tau \cdot
\vec N\left(\dsf{\vec r-\vec R}{|\vec r-\vec R|}\right)
F(\vec r-\vec R)\right]\,,
\label{ansatz}
\end{equation}
where $\vec N$ is the unit isotopic vector correlated with the unit space
vector $({\vec r-\vec R})/{|\vec r-\vec R|}$.  In the more general case, $\vec
N$ is a functional of some angular functions $\Theta(\theta)$,
$\Phi(\theta,\varphi) $~\cite{Nikolaev,Otofuji}. The dependence $\vec N=\vec
N[\Theta(\theta),\Phi(\theta,\varphi)]$ and the dependence of the profile
function $F$ on the direction of the radial co-ordinate space vector $\vec
r-\vec R$ characterizes the deformation of the skyrmion. As it was mentioned
above, in the works~\cite{Kopeliovich,Nikolaev,Otofuji} such deformation was
taken into account when considering properties of multi-skyrmionic systems.
These modifications are dictated by the topological properties of the model,
in particular the $\Phi(\theta,\varphi)$ dependence is dictated by the
constraint that the baryon number is larger than one, $B>1$.  Here we consider
the $B=1$ case and simply set $\Phi(\theta,\varphi)=\varphi$. Consequently, we
can represent the vector $\vec N$ in the following form
\begin{equation}
\vec N=\{\sin\Theta(\theta)\cos\varphi,\sin\Theta(\theta)\sin\varphi,
\cos\Theta(\theta)\}\,.
\label{greattheta}
\end{equation}
However, this is not the only deformation effect. We also modify the
profile function $F$ by allowing for a $\theta$ dependence,
\begin{equation}
F=F(|\vec r-\vec R|,\theta)\,.
\label{profform}
\end{equation}
Note that any azimuthal $\varphi$ dependence of $F$ is removed due to the
axial symmetry of the system (see the next subsection). Obviously these
deformation effects also exist in the case when a skyrmion is placed at the
center of nucleus, $R=0$, but in this case the deformation has only a radial
dependence.  In the previous work~\cite{prc001} this dependence was determined
by solving the equations for the profile function of the ansatz~\re{anzsph} in
nuclear matter, $F (r) \to F^* (r)$.  It is clear that in the present case the
deformation has in general both a radial and an angular dependence. The
angular dependence will come into play when the skyrmion is located at some
distance from the center of the nucleus.  The radial part of the deformation
represents the breathing mode of the skyrmion and the angular part represents
its deviations from the spherical form.  Finally, taking into account our
choice of $\vec N$ and $F$ (eqs.~\re{greattheta} and~\re{profform}), the
following ansatz for the chiral field $U$ arises
\begin{equation}
U(\vec r-\vec R)=\exp\left[i\vec\tau \cdot \vec N(\Theta(\theta),\varphi)
F(|\vec r-\vec R|,\theta)\right]~.
\label{ansatzlast}
\end{equation}
This form will be used in the present work.  As a result some modifications
beyond the one induced by the breathing mode will appear. For example, one
expects that an intrinsic 
quadrupole moment of the skyrmion emerges. This will be
discussed in more detail below.

Before discussing the in-medium skyrmion properties, we point out that
throughout this paper we neglect the effects due to Fermi motion of the
solitons (nucleons). In a strict large-$N_c$ counting, such effects are
suppressed. However, since we do not expect deformation effects to be large,
one can doubt the validity of this formal suppression. At present, it is
simply not known how to include such effects systematically because one not
only would we have to boost the solitons (where the boost velocity could be
obtained in the local density approximation) but also the coupling to the
rotational modes would considerably complicate the quantization procedure to
obtain physical states.  While the neglect of Fermi motion might to some
extent alter the quantitative results of our study, we believe, however, that
the qualitative ones would remain unaltered.

%%%%%%%%%%%%%%%%%%%%%%%%%%%%%%%%%% subsection 2.2
\subsection{Mass functional in the static case}

\begin{figure}[hbt] %%% figure 1
   \epsfysize=8cm
   \centerline{\epsffile{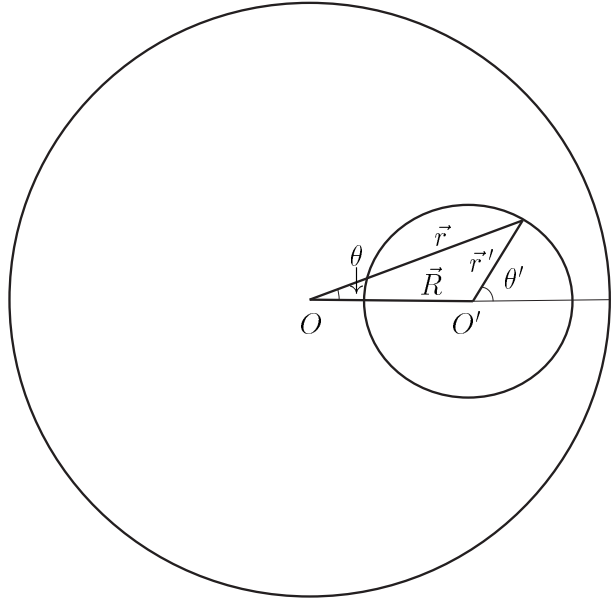}}
   \vspace{1.cm}
   \centerline{
\parbox{12cm}{
\caption{
\label{fig1}
Location of a skyrmion inside the given nucleus. $O$ is the center
of the nucleus and $O'$ is the topological center of the skyrmion.
We define by  $|\vec R|$ the distance between these centers. 
The direction of $\vec R$ coincides with 
directions of axes $\vec z$ and $\vec z\,'$.
}}}
\end{figure}

The mass functional in the static case can be obtained from the
lagrangian~\re{lag} by using the ansatz~\re{ansatzlast}
\begin{equation}
\begin{array}{l}
M(\vec R) = 
 -\ds\int {\cal L}_{\rm static}[U(\vec r-\vec R),\vec r \,]d^3\vec r\,.
\end{array}
\end{equation}
To simplify the calculations one introduces the variable $\vec r\,' = \vec r -
\vec R$ putting the origin of the coordinate system $O$ at the topological
center of the skyrmion $O'$ (see Fig.~\ref{fig1}).  Then the expression for
the mass functional can be written as
\begin{equation}
\begin{array}{l}
M(\vec R) = -\ds\int{\cal L}_{\rm static}[U(\vec r\,'),\vec r\,',\vec R]d^3\vec
r\,'\,.
\label{massr}
\end{array}
\end{equation}
We note that the distance $|\vec R|$ between the skyrmion and the nucleus is a
matter of convention. It depends on how one determines the center of the
soliton. If the skyrmion is deformed its mass and topological centers may not
coincide.  In this paper we determine $|\vec R|$ as the distance between the
center or the nucleus under consideration and topological center of the
skyrmion under consideration.

From figure~\ref{fig1} it is easy to see that the system has an axial
symmetry. One can choose the direction of the $z'$ axis along the direction of
the vector $\vec R$ so that the integrand in~eq.~\re{massr} has no azimuthal
$\varphi'$ dependence.  Consequently the profile function $F$ only depends on
the variables $r'$ and $\theta'$. Therefore, the lagrangian~\re{lag} in the
static case can be written in the form
\begin{equation}
\begin{array}{l}
{\cal L}_{\rm static}=-\dsf{F_\pi^2}{8}[(\partial_i \vec N_j 
 \cdot\partial_i \vec N_j)
\sin^2 F+\partial_i F\partial_iF]\alpha_p(r\,',\theta',R)\\
\quad \\\qquad\qquad
-\dsf{1}{4e^2}
[(\partial_i\vec N\cdot\partial_i \vec N)^2-
(\partial_i\vec N\cdot\partial_j \vec N)^2] \sin^4F\\
\quad \\\qquad\qquad
-\dsf{1}{2e^2}
[(\partial_i\vec N\cdot\partial_i \vec N)\partial_j F\partial_jF-
(\partial_i\vec N\cdot\partial_j \vec N)
\partial_i F\partial_jF] \sin^2F\\
\quad \\\qquad\qquad
-\dsf{m_\pi^2F_\pi^2}{4}(1-\cos F)\alpha_s(r\,',\theta',R)~,
\label{lstatic}
\end{array}
\end{equation}
which (after some algebraic manipulations) leads to following mass
functional\footnote{ Hereafter, we drop the prime on the variable $\theta$ for
  convenience.}
\begin{equation}
\begin{array}{l}
M(R) =\dsf{\pi F_\pi}{e}
\ds\int\limits_0^\infty\int\limits_{0}^{\pi}
 \left\{
\dsf{\sin^2\Theta}{\sin^2\theta}\dsf{\sin^4F}{\tilde r^2}\Theta_\theta^2
+\dsf{1}{4}\sin^2F
\left(\dsf{\sin^2\Theta}{\sin^2\theta}+\Theta_\theta^2\right)
\alpha_p(\tilde r,\theta,\tilde R)\right.\\
\quad \\\qquad \qquad
+\left[\dsf{\sin^2\Theta}{\sin^2\theta}\dsf{\sin^2F}{\tilde r^2}+
\dsf{\alpha_p(\tilde r,\theta,\tilde R)}{4}\right]
F_\theta^2+\dsf{m_\pi^2}{e^2F_\pi^2}\left(\sin\dsf{F}{2}\right)^2
\tilde r^2\alpha_s(\tilde r,\theta,\tilde R)\\
\quad \\\qquad \qquad
\left.
+\left[\sin^2F\left(\dsf{\sin^2\Theta}{\sin^2\theta}+\Theta_\theta^2\right)
+\dsf{\tilde r^2\alpha_p(\tilde r,\theta,\tilde R)}{4}\right]
F_{\tilde r}^2\right\}
\sin \theta d\theta d\tilde r\,.
\label{masstat}
\end{array}
\end{equation}
Here $F_{\tilde r}=\dsf{\partial F}{\partial \tilde r}$,
$F_\theta=\dsf{\partial F}{\partial \theta}$, 
$\Theta_\theta=\dsf{\partial \Theta}{\partial \theta}$
and we have introduced the dimensionless variables $\tilde r=eF_{\pi}r'$,
$\tilde R=eF_{\pi}R$.

The corresponding baryon number of the skyrmion is given by
\begin{equation}
B = -\dsf{1}{\pi}\int\limits_0^{\infty}\int\limits_{0}^{\pi}
F_{\tilde r}\Theta_\theta\sin^2F\sin \Theta  d\theta d\tilde r\,.
\label{baryonnumber}
\end{equation}

The static soliton solution is found by 
minimizing the  functional~\re{masstat}. For doing that,
one has to solve a system of coupled differential equations
\begin{equation}
\begin{array}{l} 
f(F_{\tilde r\tilde r},F_{\theta\theta},
 F_{\tilde r},F_{\theta},\Theta_\theta,F,\Theta)=0~,
\nonumber\\
\quad\\
g(\Theta_{\theta\theta},\Theta_\theta,
 F_{\tilde r},F_{\theta},\Theta,F)=0~,\nonumber
\end{array}
\end{equation}
for the functions $F(\tilde r,\theta)$ and $\Theta(\theta)$ subject to the
boundary conditions determined from the baryon number condition
$B=1$~\re{baryonnumber}.  To simplify the numerical computations we perform
various approximations. First, consider the profile function $F(r)$ for an
undeformed (spherical) skyrmion. It is well known that $F(r)$ can be
parameterized to a very good approximation as
\begin{equation}
F=2\arctan\left(\dsf{r_S^2}{r^2}\right)
\label{profpar}
\end{equation}
for the free skyrmions with baryon number $B=1$. The value of the parameter
$r_S$ is obtained from the constraint of having the minimum mass functional
for the free skyrmion. This parameterization in principle can be used for a
bound skyrmion and we apply this ansatz in constructing our profile function
(despite some short-coming, see section~\ref{sec:mass}). Also, from previous
calculations we know that the dependence of the profile function on the
density is weak, see e.g.~\cite{prc001}.  Renormalization of the mass
functional~\re{masstat} mainly occurs via the medium functionals $\alpha_p$
and $\alpha_s$.  Furthermore, from Fig.~\re{fig1} we see that changes of
$\theta'$ mainly affect the value of medium functionals\footnote{ We remind
  that $\alpha_p$ and $\alpha_s$ are functionals of the medium density
  $\rho(r)$.} and following the results from ref.~\cite{prc001}, we assume
changes of profile function to be small.  Using these facts we can represent
our profile function as
\begin{equation}
F=2\arctan\left\{\left(\dsf{r_S^2}{r^{\prime2}}\right)[1+\gamma_1\cos\theta+
\gamma_2\cos^2\theta+\gamma_3\cos^3\theta+\dots]\right\}~,
\label{proflast}
\end{equation}
where the $\cos$ functions are chosen to maintain periodicity in $\theta$.
For the function $\Theta$ we choose the parameterization 
\begin{equation}
\Theta=\theta+\delta_1\sin 2\theta+\delta_2\sin 4\theta
 +\delta_3\sin6\theta+ \dots
\label{thetalast}
\end{equation}
in order to avoid  singularities from factors like ${\sin\Theta}/{\sin\theta}$
in the expression of the mass functional~\re{masstat}\cite{Nikolaev,Otofuji}.
In eqs.~\re{proflast} and \re{thetalast},
 $r_S$, $\gamma_1$, $\gamma_2$, $\gamma_3$, $\ldots$ and 
$\delta_1$, $\delta_2$, $\delta_3$, $\ldots$ are variational (or in other
words deformation) parameters.

Our aim is the minimization of mass the functional~\re{masstat} using the
expressions ~\re{proflast} and~\re{thetalast} for the profile function. Before
doing that, we have to specify the medium dependence that enters through the
$s$- and $p$-wave pion-nucleon functionals $\chi_s$ and $\chi_p$,
respectively.

%%%%%%%%%%%%%%%%%%%%%%%%%%%%%%%%%% subsection 2.3
\subsection{Medium functionals and parameterization of the density}
\label{sec:sus}
The medium functionals $\chi_s$ and $\chi_p$
introduced in Eqs.~\re{lag},~\re{medfunc}
have the form~\cite{prc001}\footnote{We consider here symmetric nuclei
with $\rho_n=\rho_p$. For asymmetric  nuclei, additional isovector
effects will be present but are not considered here.}
\begin{eqnarray}\label{funcrho}
\chi_s(\tilde r,\theta,\tilde R) &=&-4\pi{\eta}b_0\rho(\tilde r,\theta,
\tilde R)\,\,, \nonumber \\
\chi_p(\tilde r,\theta,\tilde R) &=& \dsf{\kappa(\tilde r,\theta,
\tilde R)}{1+g_0'\kappa(\tilde r,\theta,\tilde R)}\,\,,
\quad\qquad\qquad
\kappa(\tilde r,\theta,\tilde R)=\dsf{4{\pi}c_0\rho(\tilde r,\theta,
\tilde R)}{\eta}\,\,,
\end{eqnarray}
where $\eta = 1 + m_\pi/M_N \sim 1.14$ is a
kinematical factor, $M_N = 938\,$MeV  the  mass of
the nucleon and $g_0^{\prime}=1/3$ is the
 Migdal parameter which takes into account the
 short--range pair correlations of the
 dipole centers. The empirical parameters
$b_0 = - 0.024 m_{\pi}^{-1}$, $c_0 =  0.21m_{\pi}^{-3}$~\cite{ericson}
can be taken from analyses of low energy pion-nucleus scattering data.

We choose the following parameterization for the density of the finite
spherical nuclei considered here~\cite{akhiezer}:
\begin{equation}
\begin{array}{ll}
\rho(r)=\left(\dsf{A-1}{A}\right)
\ds {\frac {2}{\pi^{3/2}r_{0}^{3}}} \left[1+\frac{Z-2}{3}
\left(\frac{r}{r_0}\right)^{2}\right]
\exp\left\{-\dsf{r^2}{r_{0}^{2}}\right\}\,,\quad &
\mbox{for $ 4 < A < 20$}\,,\\
\quad \\
\rho(r)
 =\left(\dsf{A-1}{A}\right)\dsf{\rho_0}
 {1+\exp\left\{\dsf{r-R'}{a}\right\}}\,,&
\mbox{for $ A\ge 20$}\,.
\label{density}
\end{array}
\end{equation}
In the actual calculations (minimization procedure) we have considered the
nuclei $^{12}$C, $^{16}$O, $^{40}$Ca, $^{56}$Fe, $^{198}$Au and $^{208}$Pb.
The first two represent {\it light}, the next two {\it medium-heavy} and the
last two {\it heavy} nuclei.  The parameter $r_0$ depends on the type of
nucleus considered.  Its value is $1.31$~fm for $^{12}$C and $1.76$~fm for
$^{16}$O.  $Z$ is the charge of the nucleus and $\rho_0=0.5m_\pi^{-3}$ is the
normal nuclear matter density. Furthermore, $R'=1.2A^{1/3}$~fm and
$a=0.58$~fm. $A$ is the baryon number of the nucleus and the factor
$(A-1)/{A}$ allows one to consider properties of a given nucleon which is one
of the nucleons of the nucleus. For the accuracy of our calculations, such a
simple rescaling of the nuclear density should be sufficient. The density is
pameterized as
\begin{equation}
 \rho(\tilde r,\theta,\tilde R)
 =\rho(\sqrt{\tilde r^2+2\tilde r\tilde R\cos\theta+ \tilde R^2})\,,
\end{equation}
as dictated by the axial symmetry of the system.

%%%%%%%%%%%%%%%%%%%%%%%%%%%%%%%%%% section 3
\section{Mass of the classical skyrmion and the baryon charge distribution}
\label{sec:mass}
For the input parameters of the Skyrme model we use $F_\pi=108$~MeV and
$e=5.265$ in order to reproduce the free space masses of the nucleon and the
$\Delta$ as in the seminal work~\cite{anwzb}.  One could also work with the
empirical value of $F_\pi$ and adjust $e$ to the $N\Delta$ mass splitting. In
that case, however, the nucleon mass would be too large, but relative changes
would be affected considerably less~\cite{UGM}. For definiteness, we prefer
here to work with the physical nucleon mass.

Our calculations have been performed using different sets of the variational
parameters $r_S$, $\gamma_1$, $\gamma_2$, $\gamma_3, \ldots ,\delta_1$,
$\delta_2$, $\delta_3, \ldots\;$. We represent the skyrmion size parameter
$r_S$ as
\begin{equation}
r_S=r_S^f-\delta r_S\,,
\label{delr}
\end{equation}
where $r_S^f$ is the value for the skyrmion in free space ($r_S^f =0.6$~fm).

The minimization procedure of the mass functional~\re{masstat} showed that
considering the set of parameters
\begin{equation}
\left\{\begin{array}{l}
\gamma_i
= 0, \mbox{ if } i\ge 3 ~,\\
\delta_j
=0, \mbox{ if } j\ge 2 ~, 
\end{array}\right.
\label{parset}
\end{equation}
is sufficiently accurate. The strength of the parameter $\gamma_i$ decreases
with increasing index $i$.  For example, $\gamma_3$ is one order of magnitude
smaller than $\gamma_1$ at the distances where deformation effects are
maximal. The parameter $\delta_j$ decrease very fast with increasing index
$j$. For example, $\delta_2$ is one order of magnitude smaller than $\delta_1$
and $\delta_3$ is almost zero. For controlling our results and keeping a high
accuracy, we have performed calculations with the following set of parameters:
\begin{equation}
\left\{\begin{array}{l}
\gamma_i
= 0, \mbox{ if } i > 5~,\\
\delta_j
=0, \mbox{ if } j > 5~.
\end{array}\right.
\label{parset1}
\end{equation}
For the reasons just mentioned, we only show results of $\gamma_1$, $\gamma_2$
and $\delta_1$.

\begin{figure}[htb]  %%% figure 2
   \epsfysize=21cm
   \centerline{\epsffile{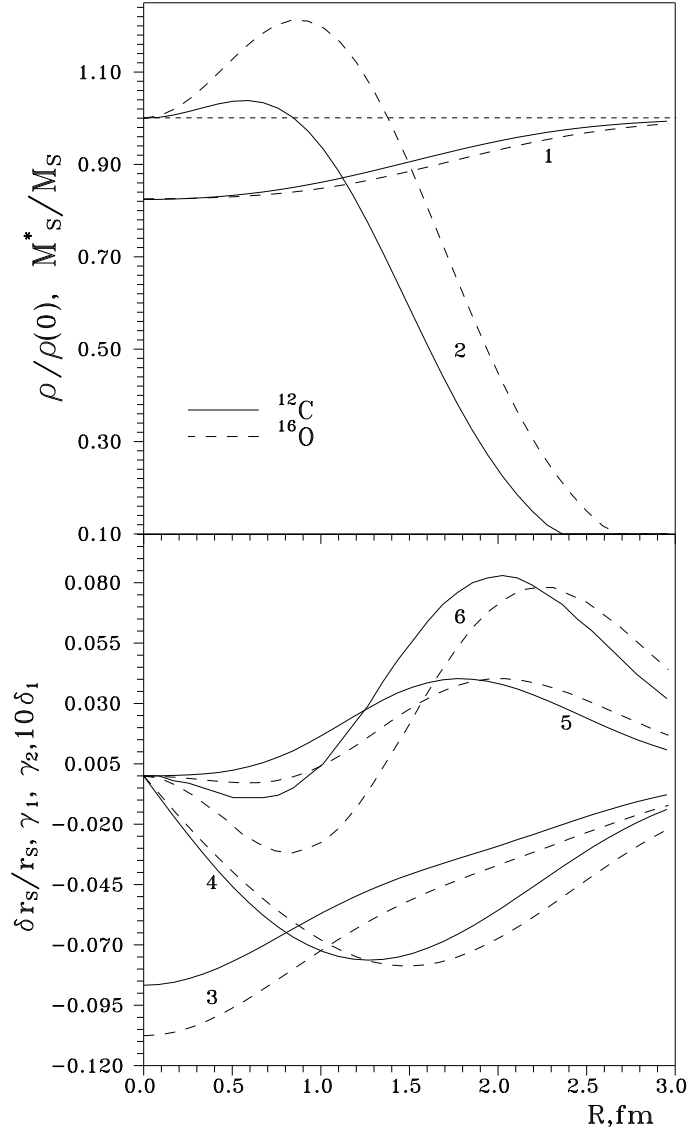}}
   \vspace{-1.5cm}
   \centerline{
\parbox{12cm}{
\caption{
\label{co}
Mass dependence of a skyrmion embedded in a given nucleus (ratio of soliton
mass in the nucleus to its free space mass, $M_S^*/M_S$) as a function of the
distance $R$ between centers of nucleus and skyrmion (band 1), the ratio of
the density at distance $R$ to its value in center of the nucleus,
$\rho/\rho(0)$ (band 2) and the dependences of the deformation parameters
$\delta r_S/r_S$ (band~3), $\gamma_1$ (band 4), $\gamma_2$ (band 5),
$\delta_1$ (band 6) on distance $R$.  Note that $\delta_1$ is multiplied by
factor 10.  Solid lines represent these dependencies for a skyrmion in
$^{12}$C and the dashed ones for that one in $^{16}$O, respectively.  We note
that the ordinate scale of the top and bottom figures are different.  For
definitions of $\delta r_S/r_S$, $\gamma_1$, $\gamma_2$ and $\delta_1$ see
eqs.~\re{proflast},~\re{thetalast} and~\re{delr}.  }}}
\end{figure}

\begin{figure}[hbt] %%% figure 3
   \epsfysize=21cm
   \centerline{\epsffile{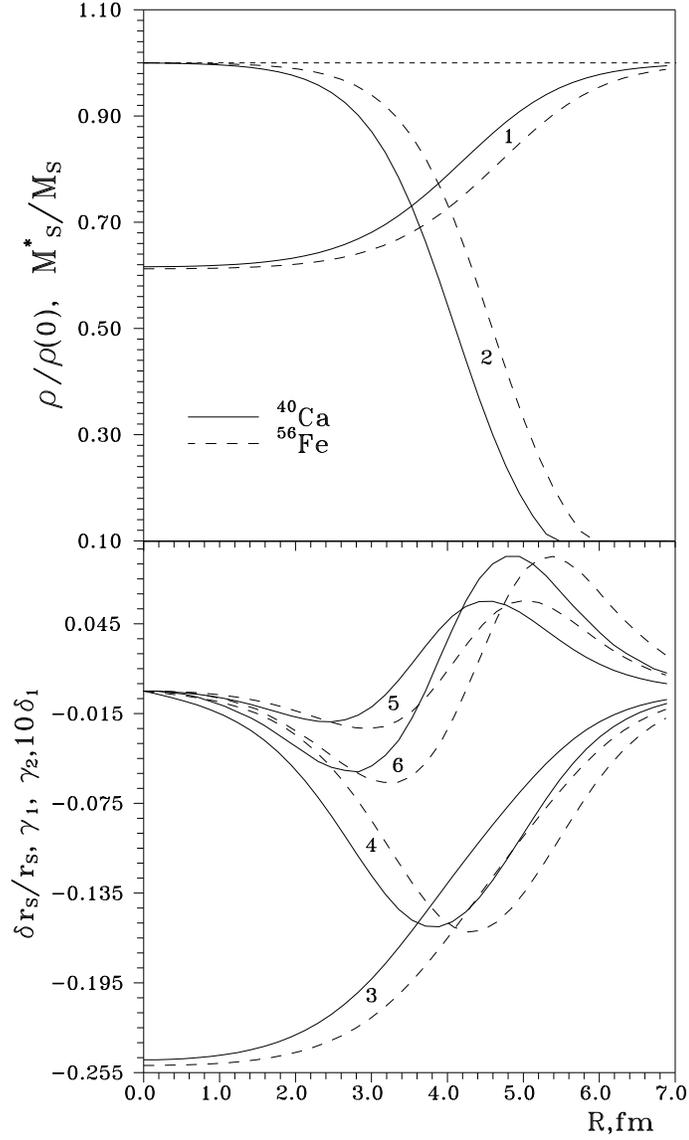}}
   \vspace{-1.5cm}
   \centerline{
\parbox{12cm}{
\caption{
\label{cafe}
As in the Fig.~\ref{co} but for a skyrmion embedded in
$^{40}$Ca (solid lines) and  $^{56}$Fe (dashed lines), 
respectively. For notations see Fig.~\ref{co}.
}}}
\end{figure}

\begin{figure}[hbt]   %%% figure 4
   \epsfysize=21cm
   \centerline{\epsffile{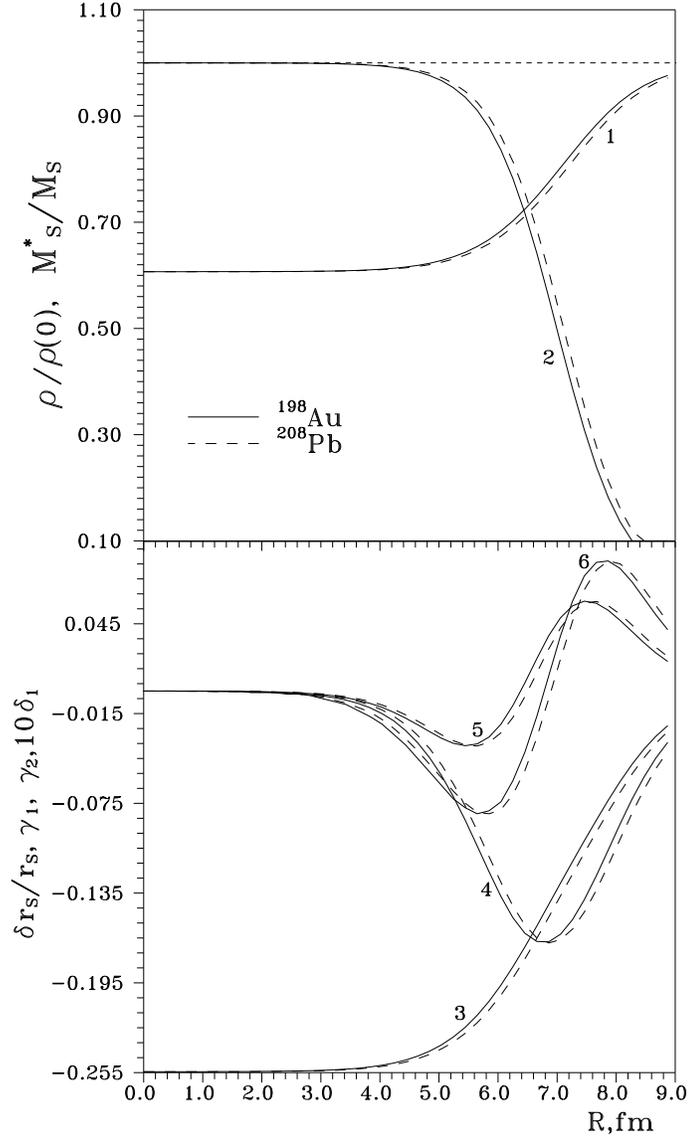}}
   \vspace{-1.5cm}
   \centerline{
\parbox{12cm}{
\caption{
\label{aupb}
As in the Fig.~\ref{co} but for a skyrmion embedded in
$^{198}$Au (solid lines) and $^{208}$Pb (dashed lines), respectively.
 For notations see Fig.~\ref{co}. }}}
\end{figure}

The dependence of the skyrmion mass from the distance between the centers of
the nucleus considered and the soliton (which represents one of the nucleons
within the nucleus) is shown in Figs.~\ref{co},~\ref{cafe},~\ref{aupb}, for
the light, medium-heavy and heavy nuclei, in order. The bands labeled ``1'' in
these figures represent the soliton mass normalized to its free space value
and for convenience we display by the bands denoted ``2'' the nuclear density
normalized to its value at the center of the nucleus.  These bands are shown
in the top panels of these figures.  The behavior of the soliton mass is
similar for all cases when the soliton is placed within various nuclei, i.e.
the mass increases with falling medium density and approaches its free space
value at the border of the nucleus. It has its minimum value at the center of
the nuclei.  This is even the case for the light nuclei, which do not have
their maximum density exactly in their center.

The deformation parameters which represent the deviations of the  nucleon
shape from the spherical one are presented in the bottom panels of
Figs.~\ref{co},~\ref{cafe},~\ref{aupb}.  The bands denoted ``3'' represent the
dependence of the relative value of the radial deformation parameter $\delta
r_S/r_S$ from soliton location inside nucleus and the bands ``4'' and ``5''
give the dependence for $\gamma_1$ and $\gamma_2$, respectively. The band
``6'' represents the parameter $\delta_1$ multiplied by factor 10. The other
deformation parameters are very small and we do not show them here. The
leading parameter $\gamma_1$ has its maximum value at the distances where the
density of the nuclei falls off by $\sim$20\% for $^{12}$C and $^{16}$O,
$\sim$ 40\% for $^{40}$Ca, $^{56}$Fe, $^{198}$Au and $^{208}$Pb, whereas the
radial deformation parameter $\delta r_S$ has its maximal value at the center
of the nuclei. The parameters $\gamma_2$ and $\delta_1$ have maximum values
where the density of the nucleus falls off $\sim$80\%. In these regions, the
density gradient is maximal leading to the largest angular deformation.  From
this one concludes that the skyrmions inside nuclei is in a deformed state.

\begin{figure}[hbt]   %%% figure 5 a and b

\vspace{-4cm}
   \epsfysize=16cm
\hspace{1.cm}\epsffile{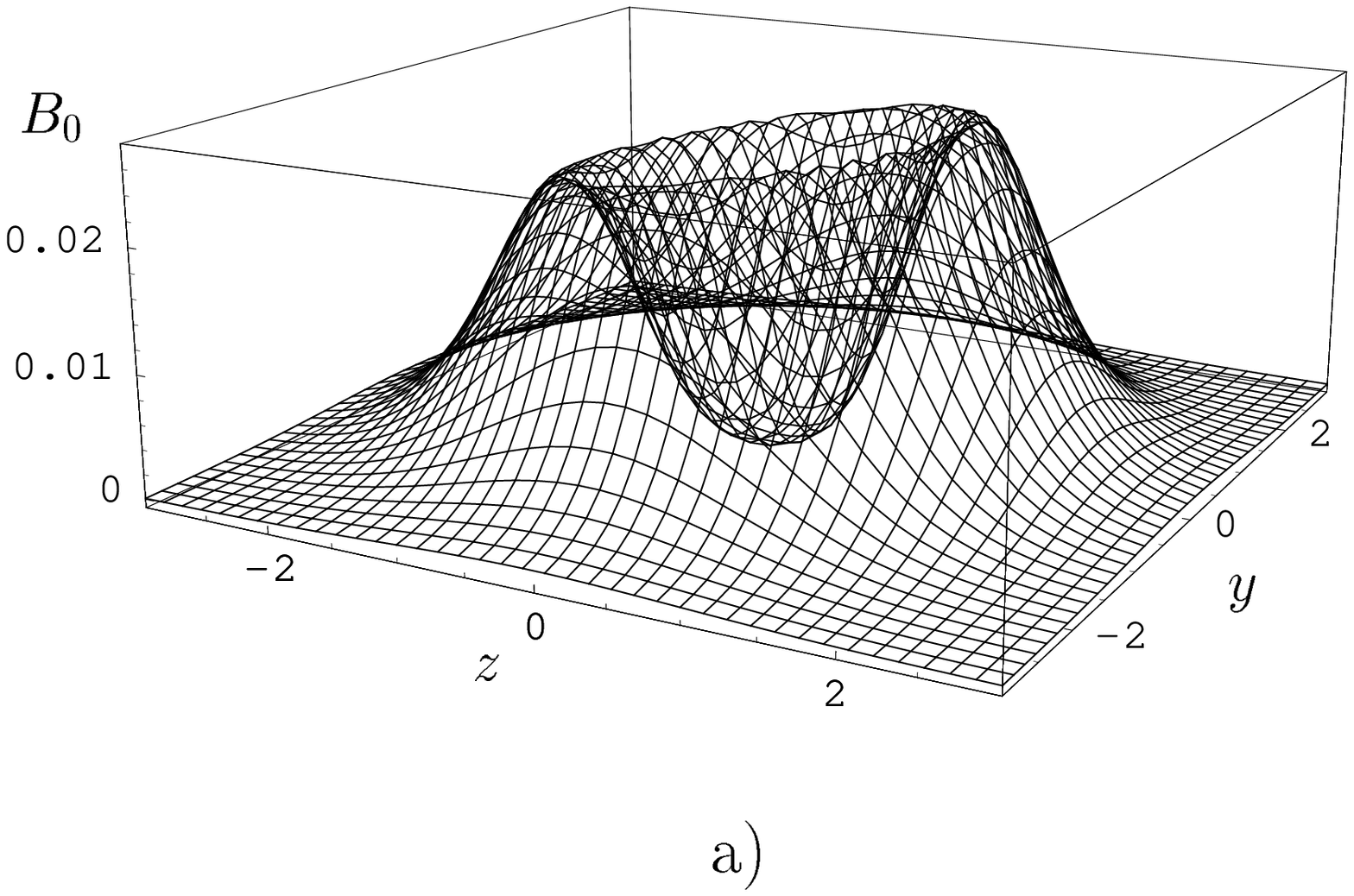}

\vspace{-8.5cm}
   \epsfysize=13cm
\centerline{\epsffile{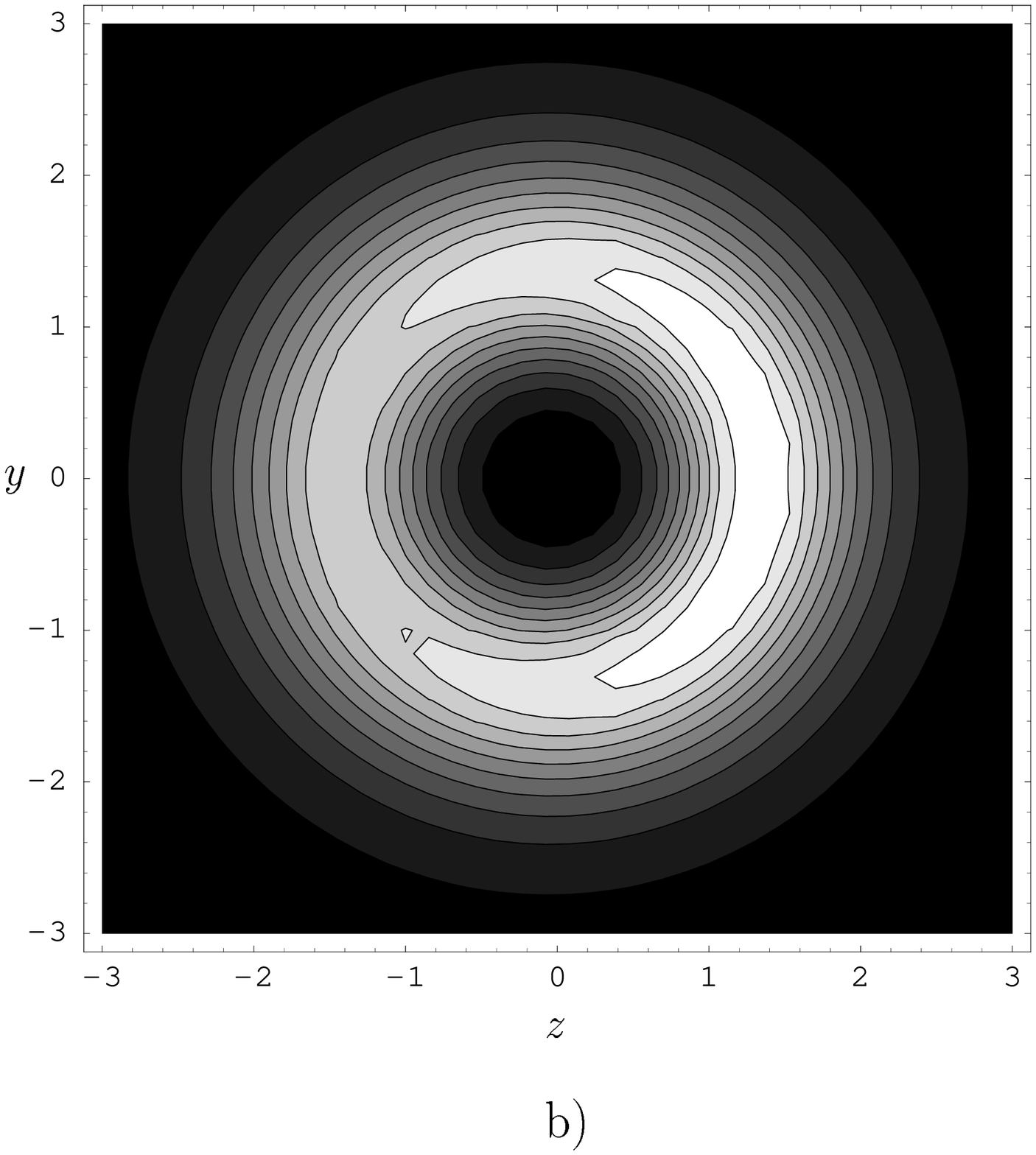}}
\vspace{-3.cm}
\begin{center}
\parbox{12cm}{
\caption{
\label{c12bc}
Baryon charge distribution $B_0$ of the skyrmion embedded in $^{12}C$ in the
$yz$ plane ({\bf a}) and its projection (equal baryon charge distribution
areas) onto this plane ({\bf b}), where $y$, $z$ are dimensionless variables.
The distance from the center of the nucleus $R$ is equal to $1.266$~fm. This
corresponds to maximum value of the leading deformation parameter $\gamma_1$.
The center of the nucleus is situated on the negative side of the axis $z$.
The strength of the baryon charge distribution corresponding to the different
area is represented by light/dark shading. The baryon density is lower if the
area is darker.  The values of the deformation parameters at this place are
$\delta r_S=-0.0476$~fm, $\gamma_1=-0.0763$, $\gamma_2=0.028$,
$\gamma_3=-0.0039$, $\delta_1=0.0029$ and $\delta_2=-0.0003$.  }}
\end{center}
\end{figure}

\begin{figure}[hbt]

\vspace{-3cm}
   \epsfysize=17cm
\hspace{1cm}\epsffile{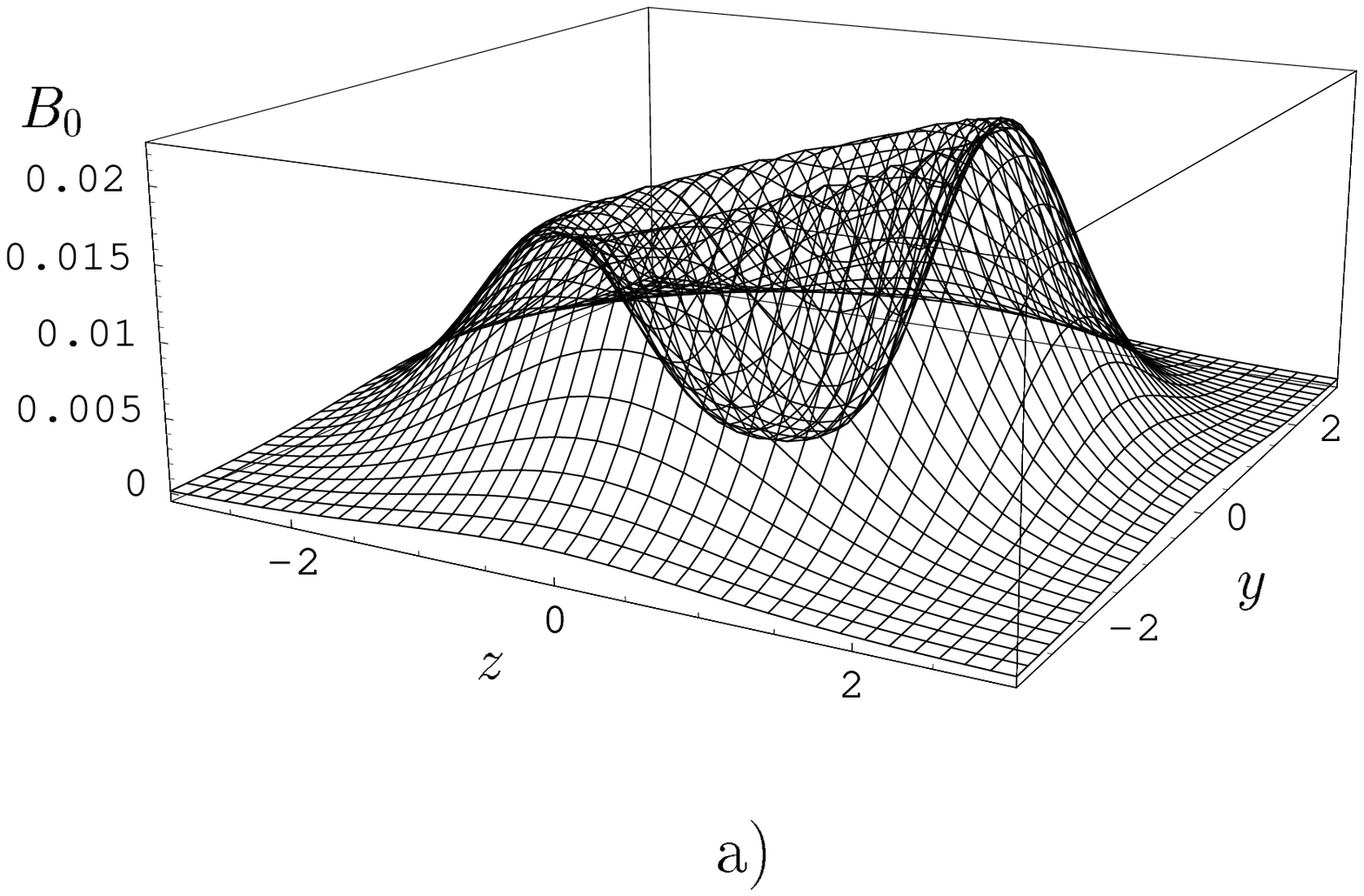}

\vspace{-8.5cm}
   \epsfysize=14cm
   \centerline{\epsffile{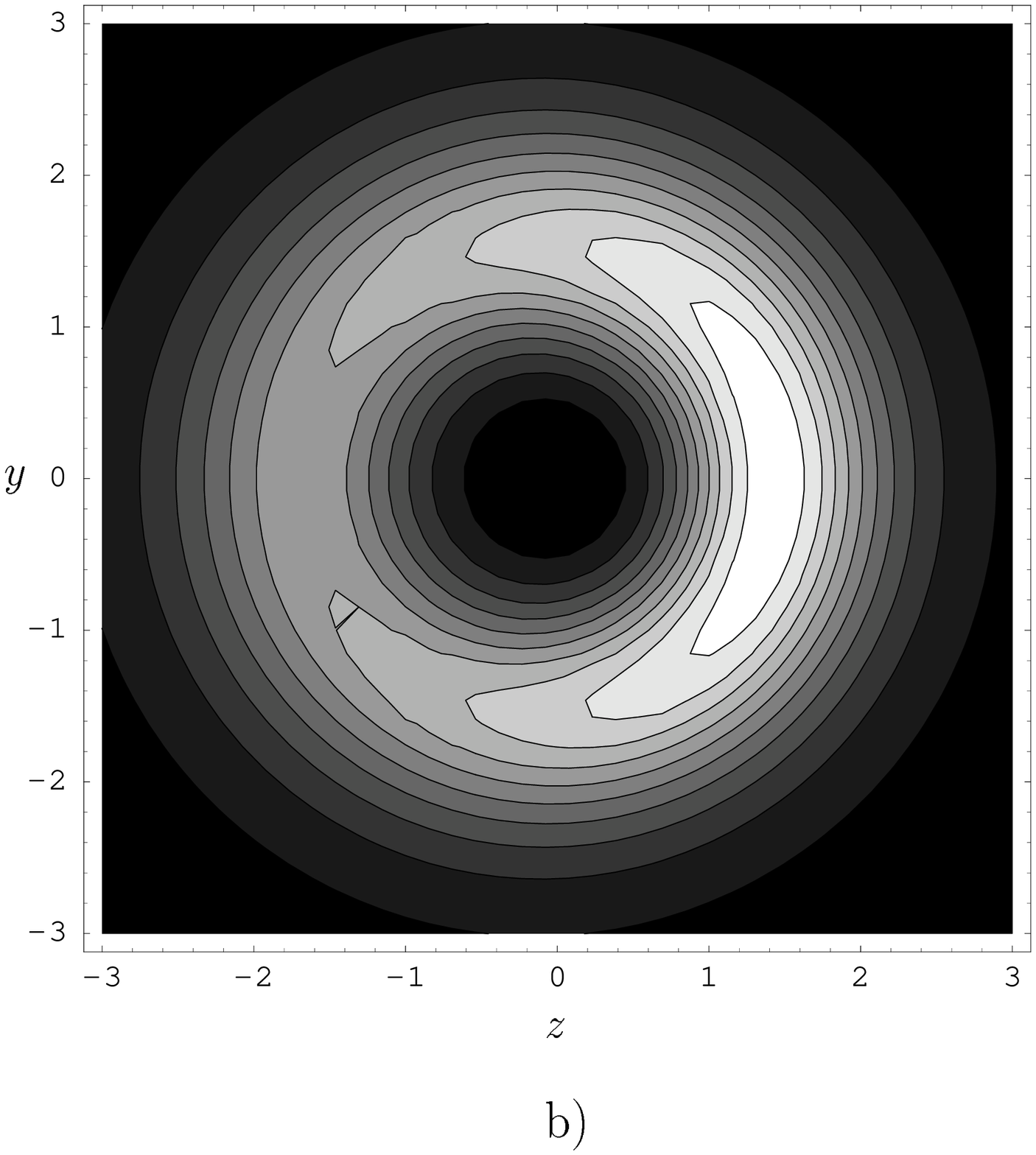}}
\vspace{-4cm}
\begin{center}
\parbox{12cm}{
\caption{
\label{Fe56bc}
As in the Fig.~\ref{c12bc} but for the skyrmion embedded in $^{56}Fe$, where
$R=4.275$~fm.  The values of the deformation parameters at this place are
$\delta r_S=-0.1464$~fm, $\gamma_1=-0.1610$, $\gamma_2=0.0326$,
$\gamma_3=0.0108$, $\delta_1=0.0026$ and $\delta_2=-0.0002$.  }}
\end{center}
\end{figure}

The topological baryon charge distribution for a skyrmion embedded in the
nuclei $^{12}$C and $^{56}$Fe is shown in Figs.~\ref{c12bc},~\ref{Fe56bc},
respectively, where figures {\bf a}) represent the baryon charge distribution
in the $y,z$ plane and figures {\bf b}) give its projection onto this plane.
Each area at the bottom figure represents the projection of surfaces of
equal baryon charge density. The point $y=z=0$ indicates the topological
center of the skyrmion and the center of nucleus is situated at the negative
side of the $z$ axis.  One can see that opposite directions in the $z$
direction are not equivalent in this plane. Stated differently, going for the
same distance in opposite directions in $z$ gives different values of the
baryon charge density. We remark that the holes at the origin of the system
($y=z=0$) are caused by the parameterization of the profile function (see
eq.~\re{proflast}). In this approximation, $F_{\tilde r}\rightarrow 0$ as
${\tilde r\rightarrow 0}$.  This derivative takes a finite value if one
directly solves the equation coming from minimization of the mass
functional~\re{masstat} not using the ansatz eq.~(\ref{profpar}), thus no such
holes would appear.  However, the effects discussed so far and in the
following are genuine and independent of the (dis)appearance of these
computational artifacts. The light area at positive values of the axis $z$ is
the area of maximal density.  The mass distribution, i.e. integrand in the
expression~\re{masstat} shows a behavior similar to the one of baryon charge
distribution.

An exotic behavior of the mass and baryon charge distributions is observed
when the skyrmion is placed at sufficiently large distances from the center of
the nucleus, where density changes are significant (see
Figs.~\ref{c12bc},~\ref{Fe56bc}).  One can see that the maximum of the baryon
density is shifted away from the center of the nuclei.  When the skyrmion is
located at the center of nucleus, the baryon density distribution is symmetric
and thus the shift becomes more effective with increasing of $R$.  The shift
of the density maximum from the topological center of the skyrmion in opposite
direction to the direction of center of the nucleus can be explained in the
following way.  The mass of skyrmion decreases in the nuclear medium and if
the density is decreasing, the skyrmion mass increases. While we are going
from the center of the nucleus to its surface the density fall-off and mass
concentration of the skyrmion occurs at the side of the smaller nuclear
density as seen from its topological center.  A similar behavior of the baryon
charge density for other nuclei is observed, so we do not present it here.

Finally, at the end of this section we remark that rotations of the figures
{\bf b)} around the $z$--axis give three dimensional surfaces with an
equal baryon charge distribution.  From this we conclude that the soliton has
two equal components of the moment of inertia which are not equal to the third
one, $I_x=I_y\ne I_z$. We will use this fact in the next section.

%%%%%%%%%%%%%%%%%%%%%%%%%%%%%%%%%% section 4
\section{Quantization procedure for non-spherical solitons}
\label{sec:quant}

As it is well known, baryons emerge in the Skyrme model as quantized
solitons~\cite{anwzb}.  This quantization is performed in terms of an
adiabatic rotation of the soliton and gives rise to the appropriate quantum
numbers for the emerging physical states. In such an approach, the grand spin
${\bf K}$, i.e. the sum of angular momentum ${\bf J}$ and isospin ${\bf T}$,
is conserved.  This leads to a tower of states with ${\bf J = T}$.  It is well
known that in this case only the nucleon ($J=T=1/2$) and the $\Delta$
($J=T=3/2)$ have physical meaning.  All other particles (higher rotational
states) should be considered artifacts of the model and will be discarded.

To quantize the solitons of our model we use the standard canonical
quantization procedure performing time--dependent rotations in isospace and in
co-ordinate space
\begin{equation}
\begin{array}{l}
U(\vec r \,')=\sigma(\vec r\,')+
i\vec\tau\cdot\vec \phi(\vec r\,')
\longrightarrow U(\vec x,t) =
\sigma(\vec x)+i\tau^i{\sf I}_{ij}(t)\phi^j(\vec x)\,,\\
\qquad \\
x_i={\sf R}^{-1}_{ij}(t)r_j'\,,
\end{array}
\label{rotansatz}
\end{equation}
where the matrix ${\sf I}_{ij}$ specifies the rotation in isospace and ${\sf
  R}_{ij}$ the one in co-ordinate space~\cite{Hajduk}. 
The parameters of the rotation
matrices ${\sf I}_{ij}$ and ${\sf R}_{ij}$ serve as collective variables
describing the rotational degrees of freedom with respect to a body-fixed
system of reference. The latter has its origin at the
topological center $O'$ of the skyrmion and points in the direction
of the distance vector $\vec R$ from the center of the nucleus to
the center of the skyrmion (see Fig.~\ref{fig1}).

The quantization procedure used here is similar to the standard
one~\cite{anwzb} but now the axial symmetry of the system should be taken into
account~\cite{Hajduk,Kopeliovich,Nikolaev,Otofuji}.  The angular velocities of
isorotations $\omega_i$ and space rotations $\Omega_i$ can be determined from
the respective equations
\begin{equation}
\dot {\sf I}_{ik}(t){\sf I}^{-1}_{kj}(t)=\epsilon_{ijk}\omega_k\,,\qquad
\dot {\sf R}^{-1}_{ik}(t){\sf R}_{kj}(t)=-\epsilon_{ijk}\Omega_k\,,
\end{equation}
where dot means differentiation with respect to the time component, and
$\epsilon_{ijk}$ is the totally antisymmetric tensor in three dimensions.

Using now the ans\"atze~\re{ansatzlast},~\re{rotansatz} in the lagrange
density~\re{lag} and integrating over the whole space one gets following
expression
%%\begin{equation}
%%\begin{array}{l}
\begin{eqnarray}
L (R)&=&\ds\int {\cal L}(R)d^3\vec r\,'=
-M(R)+(\omega_1^2+\omega_2^2)I_{\omega\omega}^{(12)}(R)+
(\Omega_1^2+\Omega_2^2)I_{\Omega\Omega}^{(12)}(R) \no\\
&& \qquad\qquad
-2(\omega_1\Omega_1+\omega_2\Omega_2)I_{\omega\Omega}^{(12)}(R)+
(\omega_3-\Omega_3)^2I_{\omega\Omega}^{(33)}(R)\,,
\label{angmom}
\end{eqnarray}
%%\begin{equation}
%%\begin{array}{l}
\begin{eqnarray}
\label{angmom1}
I_{\omega\omega}^{(12)}(R) &=& 
\dsf{\pi}{2e^3F_\pi}\ds\int\limits_0^\infty\int\limits_{0}^\pi
\sin^2F\left[(3+\cos2\Theta)\left(\dsf{F_\theta^2}{2}+\dsf{\tilde r^2}{8}
(1+4F_{\tilde r}^2)\right)\right. \no \\ %\quad \\ 
&& \qquad\qquad
\left.+\sin^2F
\left(\dsf{\sin^2\Theta}{\sin^2\theta}
 +\Theta_\theta^2\cos^2\Theta\right)\right]
\sin\theta d\theta d\tilde r\,,\\
\label{angmom2}
I_{\Omega\Omega}^{(12)}(R) &=&
\dsf{\pi}{2e^3F_\pi}\ds\int\limits_0^\infty\int\limits_{0}^\pi
\left[\dsf{\tilde r^2}{4}\dsf{\sin^2\Theta \cos^2\theta}{\sin^2\theta}\sin^2F+
\left(\dsf{\tilde r^2}{4}+\dsf{(3+\cos2\theta)}{2}
\dsf{\sin^2\Theta}{\sin^2\theta}\sin^2F\right)F_\theta^2
\right.\no \\ %%\quad\\
&& \qquad\qquad\left.
+\dsf{\tilde r}{2}\left(
\tilde r^2+4\dsf{\sin^2\Theta}{\sin^2\theta}\sin^2F\right)
F_\theta F_{\tilde r}\sin(\Theta-\theta)
\right.\no \\ %%\quad\\
&&\qquad\qquad\left.
+\dsf{\tilde r^2}{4}\left(2\,[\,2-\cos2(\Theta\!-\!\theta)+\cos2\theta\,]
\dsf{\sin^2\Theta}{\sin^2\theta}\sin^2F+\tilde r^2\sin^2(\Theta\!-\!\theta)
\right)
F_{\tilde r}^2
\right.\no \\
%%\quad\\
&&\qquad\qquad\left.
+\dsf{1}{4}\sin^2F\left(\tilde r^2
+2(3+\cos2\theta)\dsf{\sin^2\Theta}{\sin^2\theta}
\sin^2F
\right)\Theta_\theta^2
\right. \no \\ %%\quad\\
&& \qquad\qquad\left.
-\dsf{\tilde r^2}{2}\sin^2F[-3+\cos2(\Theta-\theta)\,]
F_{\tilde r}^2\Theta_\theta^2\right]
\sin\theta d\theta d\tilde r\,,\\
\label{angmom3}
%%&&\quad\\
I_{\omega\Omega}^{(12)}(R) &=& 
\dsf{\pi}{2e^3F_\pi}\ds\int\limits_0^\infty\int\limits_{0}^\pi
\sin^2F\left[
\left(\dsf{F_\theta^2}{2}+\dsf{\tilde r^2}{8}
(1+4F_{\tilde r}^2)\right)\dsf{\cos\theta}{\sin\theta}\sin2\Theta
\right. \no \\ %%\quad\\
&& \qquad\qquad\left.
+\left(\dsf{\tilde r^2}{4}+
\dsf{\sin^2\Theta}{\sin^2\theta}\sin^2F-
\tilde rF_{\tilde r}
[F_\theta\sin(\Theta-\theta)-\tilde rF_{\tilde r}]\right)\Theta_\theta
\right. \no\\ %%\quad\\
&&\qquad\qquad\left.
+\dsf{\cos\theta}{2\sin\theta}\Theta_\theta^2\sin^2F\sin2\Theta\right]
\sin\theta d\theta d\tilde r\,,\\ %%\quad\\
I_{\omega\Omega}^{(33)}(R) &=& 
\dsf{\pi}{2e^3F_\pi}\ds\int\limits_0^\infty\int\limits_{0}^\pi
\sin^2F\left(
\dsf{\tilde r^2}{2}+
2F_\theta^2+2\tilde r^2F_{\tilde r}^2
+2\Theta_\theta^2\sin^2F\right)\sin^2\Theta
\sin\theta d\theta d\tilde r\,,
\label{angmom4}
\end{eqnarray}
%\end{array}
%\end{equation}
where $M(R)$ is static mass functional (eq.~\re{masstat}) and
$I_{\omega\omega}^{(12)}$, $I_{\Omega\Omega}^{(12)}$,
$I_{\omega\Omega}^{(12)}$, $I_{\omega\Omega}^{(33)}$ are four moments of
inertia of the skyrmion.  In deriving equations~\re{angmom}-\re{angmom4} we
have made use of the axial symmetry of the system.

\noindent
Defining now canonical conjugate variables in the body-fixed reference system
\begin{equation}
T_i=\dsf{\partial {\cal L}}{\partial \omega_i} ~, \quad {\rm and} \quad
J_i=\dsf{\partial {\cal L}}{\partial \Omega_i}~,
\end{equation}
one obtains the Hamiltonian of the system
\begin{eqnarray}
%\begin{array}{l}
\hat H&=&M+\dsf{
(\hat T_1^2+\hat T_2^2)I_{\Omega\Omega}^{(12)}
+(\hat J_1^2+\hat J_2^2)I_{\omega\omega}^{(12)}
+2(\hat T_1\hat J_1
+\hat T_2\hat J_2)I_{\omega\Omega}^{(12)}
}{4\left[I_{\omega\omega}^{(12)}I_{\Omega\Omega}^{(12)}-
\left(I_{\omega\Omega}^{(12)}\right)^2\right]}+\dsf{\hat
T_3^2}{4I_{\omega\Omega}^{(33)}} \nonumber\\
%\quad\\
&=& M+\dsf{\hat {\bf T}^2I_{\Omega\Omega}^{(12)}
+\hat {\bf J}^2I_{\omega\omega}^{(12)}
+2\hat {\bf T} \cdot \hat{\bf J}I_{\omega\Omega}^{(12)}
-\left(I_{\Omega\Omega}^{(12)}
+I_{\omega\omega}^{(12)}
-2I_{\omega\Omega}^{(12)}\right)\hat T_3^2
}{4\left[I_{\omega\omega}^{(12)}I_{\Omega\Omega}^{(12)}-
\left(I_{\omega\Omega}^{(12)}\right)^2\right]}+\dsf{\hat
T_3^2}{4I_{\omega\Omega}^{(33)}}\,.
%\end{array}
\label{hamiltonian}
\end{eqnarray}
Here ${\bf \hat T}$ is the isospin operator, ${\bf \hat J}$ the spin operator.
If in addition one considers states with $K=0$, where ${\bf K}={\bf T}+{\bf
  J}$ is the grand spin, and sandwiches the Hamiltonian between wave functions
$|T,T_3;J,J_3=-T_3\rangle$ in the body-fixed system of reference, 
the energy of the system takes the form
\begin{equation}
\begin{array}{l}
E=M+\dsf{
\left[
I_{\omega\omega}^{(12)}
+I_{\Omega\Omega}^{(12)}
-2I_{\omega\Omega}^{(12)}\right]
\left(T(T+1)-T_3^2\right)
}
{4\left[I_{\omega\omega}^{(12)}I_{\Omega\Omega}^{(12)}-
\left(I_{\omega\Omega}^{(12)}\right)^2\right]}
+\dsf{T_3^2}{4I_{\omega\Omega}^{(33)}}\,,
\end{array}
\label{energy}
\end{equation}
where we used the constraint $T_3=-J_3$ which is due to the axial symmetry of
ansatz~\re{ansatzlast}.  The degeneracy on the quantum number $J_3$ is
partially lifted, i.e.\ the energy of the skyrmion only coincides for states
with $\pm J_3$.  For example, the following pairs of baryons have the same
masses:
\begin{equation}
\begin{array}{lll}
 i)\quad& |p,J_3=-\frac{1}{2}\rangle & \mbox{and}\quad 
     |n,J_3=\frac{1}{2}\rangle\,;\\
 ii)& |\Delta^+,J_3=-\frac{1}{2}\rangle & \mbox{and}\quad 
     |\Delta^0,J_3=\frac{1}{2}\rangle\,;\\
 iii)& |\Delta^{++},J_3=-\frac{3}{2}\rangle & \mbox{and}\quad 
     |\Delta^-,J_3=\frac{3}{2}\rangle\,,
\label{states}
\end{array}
\end{equation}
but the baryons in the states $ii)$ and $iii)$ do {\it not} have the same
masses. The above specified quantum numbers refer to the body-fixed
system of reference. In contrast to a deformed skyrmion in an isotropic
background (e.g.\ the vacuum or  isotropic and isosymmetric 
nuclear matter) \cite{Hajduk},
our present case corresponds to a deformed nucleon or
delta  in an unisotropic background given by the nuclear density profile. 
Therefore, the projection onto good quantum numbers in the laboratory
system of reference does not follow here from a simple rotation of
the intrinsically deformed object. In fact, only the third components
of angular momentum and isospin in the body-fixed system, $J_3$ and 
$T_3$, are good quantum numbers. 

As a check, let us consider whether we can obtain the canonical spherically
symmetric solutions from our deformed ones. First, it is easy to see that
\begin{equation}
\lim_{\begin{array}{l}
{F_\theta\rightarrow 0}\\
{\Theta\rightarrow \theta}
\end{array}} I_{\omega\omega}^{(12)} =
\lim_{\begin{array}{l}
{F_\theta\rightarrow 0}\\
{\Theta\rightarrow \theta}
\end{array}} I_{\Omega\Omega}^{(12)} =
\lim_{\begin{array}{l}
{F_\theta\rightarrow 0}\\
{\Theta\rightarrow \theta}
\end{array}} I_{\omega\Omega}^{(12)} =
\lim_{\begin{array}{l}
{F_\theta\rightarrow 0}\\
{\Theta\rightarrow \theta}
\end{array}} I_{\omega\Omega}^{(33)} =\dsf{I}{2}~,
\end{equation}
where $I$ is the  moment of inertia  of the spherically symmetric hedgehog
ansatz,
\begin{equation}
I = \dsf{2\pi}{3e^3F_\pi}\ds\int\limits_0^\infty \sin^2F\left(
\tilde r^2+
4\tilde r^2F_{\tilde r}^2+4\sin^2F\right)\, d\tilde r\,.
\end{equation}
Consequently
\begin{equation}
\lim_{\begin{array}{l}
{F_\theta\rightarrow 0}\\
{\Theta\rightarrow \theta}
\end{array}} 
\dsf{I_{\omega\omega}^{(12)}I_{\Omega\Omega}^{(12)}-
\left(I_{\omega\Omega}^{(12)}\right)^2}{
 I_{\omega\omega}^{(12)}+I_{\Omega\Omega}^{(12)}-
2I_{\omega\Omega}^{(12)}}=
\lim_{\begin{array}{l}
{F_\theta\rightarrow 0}\\
{\Theta\rightarrow \theta}
\end{array}} 
\dsf{\left(I_{\omega\omega}^{(12)}-I_{\omega\Omega}^{(12)}\right)
I_{\omega\Omega}^{(12)}}{I_{\omega\omega}^{(12)}-I_{\omega\Omega}^{(12)}}=
\lim_{\begin{array}{l}
{F_\theta\rightarrow 0}\\
{\Theta\rightarrow \theta}
\end{array}} 
I_{\omega\Omega}^{(12)}~,
\end{equation}
and one obtains the well known energy expression for a baryon in the
spherically symmetric case
\begin{equation}
E=M+\dsf{T(T+1)}{2I}~.
\end{equation}
More generally, the following relation for all observables can be shown to
hold:
\begin{equation}
\lim_{\begin{array}{l}
{F_\theta\rightarrow 0}\\
{\Theta\rightarrow \theta}
\end{array}} 
\left(\begin{array}{c}
\mbox{Expression in axially}\\
\mbox{symmetric case}
\end{array}\right)=
\left(\begin{array}{c}
\mbox{Expression in spherically}\\
\mbox{symmetric case}
\end{array}\right)\,.
\end{equation}
This shows that our construction has the proper limit when all
deformation parameters are set to zero.

%%%%%%%%%%%%%%%%%%%%%%%%%%%%%%%%%% section 5
\section{Quantum properties of solitons}
\label{sec:prop} 
%%%%%%%%%%%%%%%%%%%%%%%%%%%%%%%%%% section 5.1
\subsection{Masses of quantized solitons}

\begin{figure}[hbt]
   \vspace{2cm}
   \epsfysize=13cm
   \centerline{\epsffile{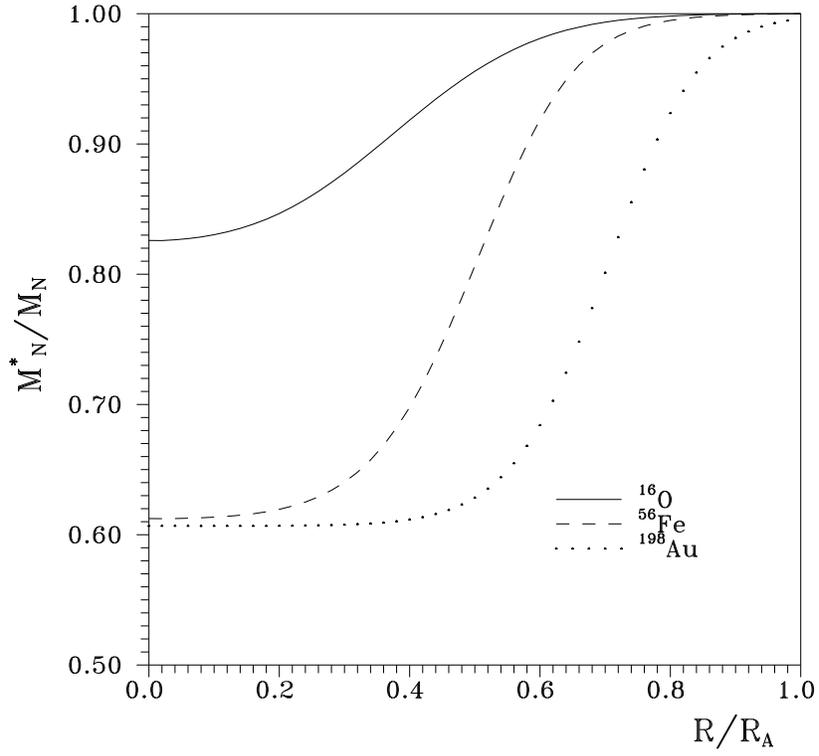}}
\vspace{2.cm}
   \centerline{
\parbox{12cm}{
\caption{
\label{nmass}
Mass dependence of a nucleon embedded in various nuclei
 as a function of the fraction $R/R_{A}$, where $R$ is the distance
 between the centers of nucleus and nucleon and $R_{A}$ is the radius
of the nucleus (see eq.~\re{rnuclei}). The solid line represents
this dependence for the nucleon in $^{16}$O, while the dashed and
dotted lines represent this dependence for $^{56}$Fe and $^{198}$Au, 
respectively. 
}}}
\end{figure}

Predictions for the masses of the quantized solitons are presented in
Fig.~\ref{nmass}, where we show the nucleon mass normalized to its free space
value. Note that the distance $R$ between the centers of the nucleon and the
nucleus is normalized to the corresponding radius $R_A$ of the nucleus for
given baryon number $A$.  We show the behavior of the nucleon mass only for
one representative nucleus of the three classes, i.e. for $^{16}$O, $^{56}$Fe
and $^{198}$Au. The other nuclei from the same class exhibit a similar
behavior.  For the radii of the nuclei we use formula
\begin{equation}
R_A=\langle R^2 \rangle^{1/2}=\dsf{\ds\int R^2\rho(R) d^3R}
 {\ds\int \rho(R) d^3R}\,.
\label{rnuclei}
\end{equation}
The behavior of the nucleon mass in the nucleus is similar for all light,
medium--heavy and heavy nuclei. When the density decreases, the mass of the
nucleon increases approaching its free space value at the edge of the nucleus.
We conclude that the mean effective mass of the nucleon in real nuclei will
not decrease as much as~$\sim 40\%$ in contrast to the case of homogeneous
nuclear matter with normal density $\rho=\rho_0$.  The finite nuclei approach
for the medium essentially improves the results, especially for light nuclei,
i.e. taking into account realistic mass distributions one obtains more
reasonable values for the mass renormalization. For example, this
renormalization is about~$\sim 17\%$ for a nucleon placed in the center of
$^{12}$C or $^{16}$O.  In principle, one could now calculate an average medium
modification in a given nucleus by simply folding the calculated effective
mass with the nuclear density. We do not show the results of such calculations
here because the calculation could be further refined.  If one uses e.g. the
shell structure of the nuclei, then all nucleons will be situated at some
distances (within the shells) from their mass center and the mean effective
value of the nucleon mass in such systems will be further reduced for nucleons
in heavy nuclei.  Further improvements of the results can be made by including
a dilaton field~\cite{joe,npa001,raku}. We also note that a similar
renormalization of the nucleon mass was found in Ref.\cite{UGM}.

As it was noted in the previous section, the mass of the $\Delta$ states
depends on the absolute value of the third component of isospin.  We found
small deviation from degeneracy for delta states with $J_3=-1/2$ and
$J_3=-3/2$\footnote{We remark that due to the neglect of the Fermi motion,
  these small effects should be considered indicative.}.  The difference of
the masses $\Delta M_\Delta=M_{\Delta^+}(J_3=-1/2)-M_{\Delta^{++}}(J_3=-3/2)$
of $\Delta$ in the states $ii)$ and $iii)$ (see eq.~\re{states}) has a
negative value near the center of the nucleus and after reaching its minimum
begins to increase towards positive values reaching at some distance its
maximal value and then smoothly drops to zero.  For example, it has its
minimal value $\Delta M_\Delta=-0.28$MeV at $R=0.84$~fm and maximal value
$\Delta M_\Delta=0.916$MeV at $R=2.447$~fm for deltas in $^{12}$C. For the
case of $^{56}$Fe, the corresponding values are $\Delta
M_\Delta$($R=4.098$~fm$)=-0.487\,$MeV and $\Delta
M_\Delta$($R=7.806$~fm$)=0.555\,$MeV. We remind the reader that all
these numbers refer to states defined in the body-fixed system of reference
as discussed before.

For the relative mass difference $M_\Delta^*-M_N^*$ we obtained a behavior
similar to the one of nucleon mass (cf. Fig.\re{nmass}) and do not graphically
present it here. The difference $M_\Delta^*-M_N^*$ decreases when the density
increases and has its minimal value at the center of the nucleus. More
specifically, $(M_\Delta^*-M_N^*)/(M_\Delta-M_N)$ has the value 0.867 for
$^{12}$C, 0.615 for $^{56}$Fe and 0.609 for $^{198}$Au, in the respective
center of the nucleus.  Note that in free space $M_\Delta-M_N=300\,$MeV in the
current approach. As to the ratio of $\Delta$ and $N$ masses,
$M_\Delta^*/M_N^*$, it is almost constant inside the nucleus and approximately
equal to that ratio in free space.

%%%%%%%%%%%%%%%%%%%%%%%%%%%%%%%%%% section 5.2
\subsection{Radii and intrinsic quadrupole moments of nucleons}

The calculation of the electromagnetic currents reduces to the calculation of
the isoscalar and the isovector currents. In topological models and in
particular in the Skyrme model, the isoscalar current is equal to the
model--independent baryon number current (for a discussion on the role of
vector mesons, see Ref.\cite{UGMrev})
\begin{equation}
B^{\mu}=\dsf{1}{24\pi^2}\epsilon^{\mu\nu\alpha\beta}
\Tr L_{\nu}L_{\alpha}L_{\beta}\,,
\label{barcur}
\end{equation}
with $\epsilon^{\mu\nu\alpha\beta}$ the totally antisymmetric tensor in four
dimensions.  Similarly, the isovector current, which is the Noether current
associated to the global transformations $U(x)\rightarrow
\exp(iQ_L)U(x)\exp(-iQ_R)$ ($Q_{L,R}$ are $2\times 2$ matrixes), has the
form\footnote{There is no summation over $\mu$ in the equation~\re{veccur}
  while there is one over the index $\nu$.}
\begin{equation}
\vec V_{\mu}=-i\dsf{F_{\pi}^2}{16} C_{\mu}\Tr\,
\vec\tau(L_{\mu}+R_{\mu})+
\dsf{i}{16{e}^{2}}\Tr\,\vec\tau\{[L_{\nu},[L_{\mu},L_{\nu}]]+
[R_{\nu},[R_{\mu},R_{\nu}]]\}\,,
\label{veccur}
\end{equation}
where
\[\quad R_{\mu}=U\partial_{\mu}U^+;\quad
C_{\mu}=\left\{\begin{array}{ccl}
1 &, &\mbox{$\mu=0$}\\
\alpha_p&, & \mbox{$\mu=1,2,3$}\\
\end{array}\right.\,. \]
Then, the operator of the electric charge of the nucleon is determined by the
zeroth components of the isoscalar and the isovector currents,
\begin{equation}
J_0^{\rm em}(\vec r \,) 
= V_0^{(3)}(\vec r\,)\hat T_3+\dsf{1}{2}B_0(\vec r\,)\,,
\end{equation}
where $V_0^{(3)}$ is the third component of the isovector current and $\hat
T_3$ is the third component of the isotopic spin operator.  Sandwiching this
current between corresponding nucleon states, we obtain the proton and neutron
charge distributions.

The normalized isoscalar root-mean-square radius along each axis,
${\langle}r_\alpha^{\prime 2}{\rangle}_{I=0}$, (normalized with the respect to
the isoscalar charge) can be determined from the zeroth component of the
baryon current and has the form
\begin{equation}
{\langle}r_\alpha^{\prime 2}{\rangle}_{I=0}=
\dsf{\ds\int\limits_0^{\infty}\int\limits_{0}^{\pi}
D_{\alpha}\tilde r^2F_{\tilde r}\Theta_\theta\sin^2F\sin\Theta 
d \theta d\tilde r}{\ds 2e^2F_\pi^2
\int\limits_0^{\infty}\int\limits_{0}^{\pi}
F_{\tilde r}\Theta_\theta\sin^2F\sin\Theta
d\theta d\tilde r} 
\,,
\qquad
 D_{\alpha}=\left\{
\begin{array}{ll}
\sin^2\theta,&\quad \alpha=1,2~,\\
2\cos^2\theta,&\quad \alpha=3~,
\end{array}\right.
\label{isosr}
\end{equation}
where $\alpha$ ($\alpha=x,y,z$) indicates the projection of the radius vector
on the corresponding axis.  Similarly, the isovector root-mean-square radii
(normalized to the isovector charge), ${\langle}r_\alpha^2{\rangle}_{I=1}$,
are determined by the zeroth component of the vector current $V_0^{(3)}$
\begin{equation}
{\langle}r_\alpha^{\prime 2}{\rangle}_{I=1}=
\dsf{
\ds\int\limits_0^{\infty}\int\limits_{0}^{\pi}D_\alpha \tilde r^2 f
\sin\theta d\theta d\tilde r}{2e^2F_\pi^2
\ds\int\limits_0^{\infty}\int\limits_{0}^{\pi}f 
 \sin\theta d\theta d\tilde r}\,,
\label{isovr}
\end{equation}
where
\begin{equation}
 f=(\sin F)^2[\tilde r^2+4F_\theta^2+4\tilde r^2F_{\tilde r}^2
 +4\Theta_\theta^2\sin^2F]\,.
\end{equation}
One can see that the influence of the medium enters indirectly via the
modified profile functions.  Results for the in--medium radii are collected in
Table~\ref{table}.  We give the isoscalar $\langle r^2\rangle^{1/2}_{I=0}$ and
isovector $\langle r^2\rangle^{1/2}_{I=1}$ root-mean-square radii of the
nucleons along the $y$ and $z$ axes inside the nuclei $^{16}$O, $^{56}$Fe and
$^{198}$Au, evaluated in the center of each nucleus and at those distances
where the intrinsic 
quadrupole moment of the nucleons has its first and second extremum,
correspondingly (see the discussion below).  We note that due to the axial
symmetry $\langle r_x^2\rangle^{1/2}_{I=0;1}= \langle
r_y^2\rangle^{1/2}_{I=0;1}$. Clearly, these radii swell, as first pointed out
in ref.\cite{noble} in the analysis of data on quasi-elastic electron-nucleus
scattering. However, this swelling is not isotropic and depends very
sensitively on the distance from the center of the nucleus. The concept of a
common swelling by one scale factor is certainly much too simple. Our results
are therefore not in contradiction with the y-scaling analysis of
ref.\cite{sick} (see also the discussion in \cite{UGM} on this issue).

\begin{figure}[hbt]
   \vspace{2cm}
   \epsfysize=21cm
   \centerline{\epsffile{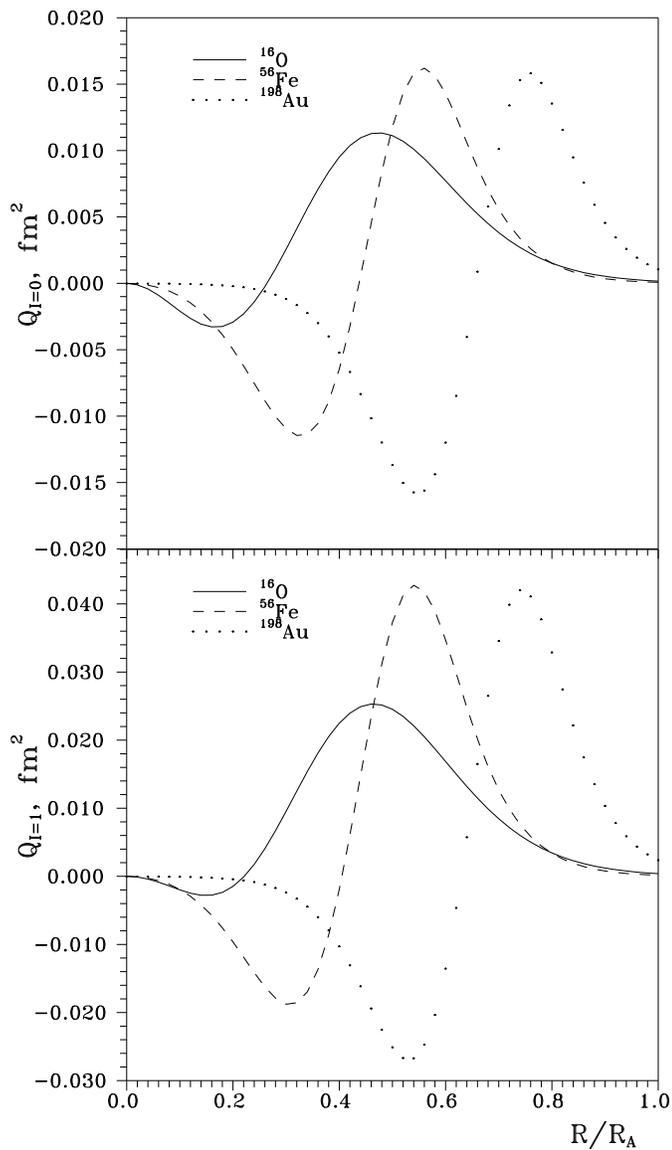}}
\vspace{0.5cm}
   \centerline{
\parbox{12cm}{
\caption{
\label{quad}
The $R/R_{A}$ dependence of isoscalar (top figure) and isovector (bottom
figure) intrinsic 
quadrupole moments of a nucleon embedded in various nuclei. For the
notations see Fig.~\ref{nmass}.  }}}
\end{figure}

As noted before due to the  deformation the nucleon can have a nonvanishing
{\em intrinsic} 
quadrupole moment~\cite{AageBen}
\begin{equation}
 Q^{\rm int}\equiv \langle J_3 |\int J_0^{\rm em}(\vec {\tilde r}) \,
\tilde r^2(3\cos^2\theta -1)\, d^3\vec{\tilde r} |J_3\rangle  
\end{equation}
that is characterized by the third component $J_3$ of the angular momentum
in the body-fixed frame of reference.
The corresponding isoscalar and isovector intrinsic 
quadrupole
moments follow from \re{isosr} and \re{isovr} by appropriate projection
\begin{equation}
\begin{array}{l}
Q_{I=0}^{\rm int}(R)=\dsf{\ds\int\limits_0^{\infty}\int\limits_{0}^{\pi} 
(3D_3-2)\tilde r^2F_{\tilde r}
\Theta_\theta\sin^2F\sin\Theta 
d\theta d\tilde r}{2 e^2F_\pi^2
\ds\int\limits_0^{\infty}\int\limits_{0}^{\pi}F_{\tilde r}
\Theta_\theta\sin^2F\sin\Theta 
d\theta d\tilde r}\,,\\
\quad\\
Q_{I=1}^{\rm int}(R)=
\dsf{
\ds\int\limits_0^{\infty}\int\limits_{0}^{\pi}(3D_3-2)\tilde r^2 f
\sin\theta d\theta d\tilde r}{2e^2F_\pi^2
\ds\int\limits_0^{\infty}\int\limits_{0}^{\pi}f\sin\theta d\theta d\tilde r}~.
\end{array}
\end{equation}
Consequently,  the proton and neutron intrinsic 
quadrupole moments follow from
\begin{equation}
Q_p^{\rm int}=\dsf{1}{2}(Q_{I=0}^{\rm int}+Q_{I=1}^{\rm int})\,,\qquad\qquad
 Q_n^{\rm int}=\dsf{1}{2}(Q_{I=0}^{\rm int}-Q_{I=1}^{\rm int})\,.
\end{equation}
The dependence of the isoscalar and isovector intrinsic 
quadrupole moments as a function
of $R/R_A$ is shown in Fig.~\ref{quad}.  Again we present results only for the
nuclei $^{16}$O, $^{56}$Fe and $^{198}$Ag. For the other nuclei the
dependences are similar. One can see that intrinsic 
quadrupole moments are small.  This
smallness of the  intrinsic 
quadrupole moments means that nucleons in the nucleus are
weakly deformed. The intrinsic 
quadrupole moments has two extrema. Near the center of
nucleus they have negative values and a corresponding minimum at that
distances where density changes are $\sim$20\% for $^{16}$O and $\sim$7\% for
$^{56}$Fe and $^{198}$Au, respectively. Then with increasing density the
intrinsic 
quadrupole moments become positive and have a maximum at the distance where
the density of the nucleus has fallen by $\sim65\div70$\%. Note also that the
isoscalar intrinsic 
quadrupole moment is about two times smaller than the isovector one.
Consequently near the center of the nucleus protons are flattened along the
axis $z$ (oblate) and and near the border of the nucleus they are extended
(prolate).  In contrast neutrons near the center of nucleus slightly prolate
and near the border of nucleus oblate.  Consequently, at some intermediate
distance the nucleons must have spherical form.

Finally, let us recall that the observable static quadrupole moment in the
laboratory system of 
an intrinsically deformed baryon in an isotropic background 
is linked to its intrinsic
quadrupole moment by the Wigner-Eckhardt theorem~\footnote{The 
first Clebsch-Gordon coefficient refers to the body-fixed frame,
while the second one refers to the laboratory frame.}
 as
follows~\cite{AageBen}
\begin{eqnarray}
 Q^{\rm lab} 
&=& \langle J\, J_3\, 2\, 0 \,|\, J J_3\rangle \langle J \pm J\, 2 \,0 
\,|\, J \pm J\rangle\, Q^{\rm int}
\nonumber
\\
&=&
\dsf{3J_3^2-J(J+1)}{(J+1)(2J+3)}\, Q^{\rm int} 
=\dsf{J(2J-1)}{(J+1)(2J+3)}\,Q^{\rm int}\,, 
\end{eqnarray}
where the relations are valid 
for maximally projected states with $|J_3|=J$.
Thus for a deformed baryon of total angular momentum $J=1/2$ 
the quadrupole moment in the {\em laboratory} system is zero for
its ground state, i.e.\ for the nucleon.
It would be interesting to investigate the transition 
quadrupole moments to excited states, extending the work
of \cite{Hajduk}.
As mentioned before, our case corresponds to a deformed baryon in
an unisotropic background. A simple projection to
the laboratory system is therefore not possible.

%%%%%%%%%%%%%%%%%%%%%%%%%%%%%%%%%% section 7
\section{Conclusion}
\label{sec:concl}
In summary, we have considered properties of a nucleon embedded in finite
nuclei making use of the Skyrme model.  Deformation effects are taken into
account by the distortion of chiral profile function under the action of the
nuclear medium, parameterized in terms of the functionals Eq.\re{funcrho}.
Nucleons and deltas emerge as quantized solitons after performing an adiabatic
rotation in co-ordinate and isospace which takes into account the axial
symmetry of the system (leading to two different moments of inertia, $I_z \neq
I_x = I_y)$.

The main results of this study can be summarized as follows:
\begin{itemize}
\item[(i)] The density dependence of the nucleon mass shows a more realistic
  behavior than in the case of a uniform density as e.g. for homogeneous
  nuclear matter. The effective nucleon mass has its minimum at the center of
  the nucleus and approaches its free space value at the surface of the
  nucleus.
\item[(ii)] The nucleons in finite nuclei are weakly deformed. They acquire a
  small intrinsic quadrupole 
   moment which, however, is strongly dependent on the
  distance from the center of the nucleus, e.g. the proton (neutron)
  deformation changes from a oblate (prolate) to a prolate (oblate) form as it
  is moved toward the surface of the nucleus.  
\item[(iii)] Similarly, there is a direction-dependent swelling of the
  isoscalar and isovector root-mean-square radii. We have stressed that the
  concept of a uniform swelling factor is too simple a concept to apply to
  real nuclei.
\item[(iv)] As consequence of the axial symmetry of the system, the $|J_3|
  =1/2$ $\Delta$ states ($\Delta^{0,+}$) and the $|J_3| =3/2$ ones
  ($\Delta^{-,++}$) have slightly different masses in finite nuclei.
\end{itemize}

Let us briefly give an outlook for further studies based on these results. As
it was mentioned above this approach can be used for model investigations of
the shell structure of nuclei and to study the properties of a remote nucleon
in such a system (halo nuclei). In the latter case the nucleon in the finite
nucleus should not be considered as one within the nucleus but weakly coupled
to the nucleus at large distances.  Also, the model naturally allows to study
nucleon properties in the lightest nuclei, especially the electromagnetic
structure of nucleons in $^3$He, which is of current experimental interest.
Work along these lines is under way.

\section*{Acknowledgments}

The work of U. Yakhshiev has been supported by INTAS YS fellowship
$N^0$00-51. We thank Andy Jackson for a useful correspondence.

%%%%%%%%%%%%%%%%%%%%%%%%%%%%%%%%%%%%%%%%%%%%%%%
\newpage

\vspace{2cm}

\begin{table}
\begin{tabular}{ccccccc}
Element&$\langle r_y^2\rangle^{1/2}_{I=0}$\,[fm]&
$\langle r_z^2\rangle^{1/2}_{I=0}$\,[fm]&
$\langle r_y^2\rangle^{1/2}_{I=1}$\,[fm]&
$\langle r_z^2\rangle^{1/2}_{I=1}$\,[fm]&
$R$\,[fm]&$\rho(R)/\rho(0)$\\
\hline
\hline
&\multicolumn{6}{c}{Center of the nucleus}\\
\cline{2-7}
$^{16}$O&0.437&0.437&0.699&0.699&0&1\\
$^{56}$Fe&0.520&0.520&0.811&0.811&0&1\\
$^{198}$Au&0.523&0.523&0.815&0.815&0&1\\
\cline{2-7}
&\multicolumn{6}{c}{First extremum}\\
\cline{2-7}
$^{16}$O&0.427&0.425&0.684&0.682&0.74&1.20\\
$^{56}$Fe&0.499&0.493&0.784&0.778&2.97&0.93\\
$^{198}$Au&0.503&0.495&0.788&0.780&5.45&0.93\\
\cline{2-7}
&\multicolumn{6}{c}{Second extremum}\\
\cline{2-7}
$^{16}$O&0.404&0.411&0.651&0.661&2.13&0.35\\
$^{56}$Fe&0.433&0.442&0.693&0.708&5.02&0.32\\
$^{198}$Au&0.435&0.444&0.696&0.711&7.46&0.31\\
\cline{2-7}
%&&&&&\\
&\multicolumn{6}{c}{Free space}\\
\cline{2-7}
   --- &0.391&0.391&0.630&0.630&---&---\\
\end{tabular}
\caption{
\label{table}
The isoscalar $\langle r^2\rangle^{1/2}_{I=0}$
and isovector $\langle r^2\rangle^{1/2}_{I=1}$ nucleon root-mean-square radii
along the  $y$ and $z$ axes inside various nuclei.
The values of these quantities are given for the center of the nuclei and
at those distances where the nucleon intrinsic 
quadrupole moments have their first
and second extremum values, correspondingly.
$R$ is the distance between centers of the nucleus and skyrmion topological 
center and the last column is the ratio of the density 
of nucleus $\rho(R)$ to its value at the center $\rho(0)$.
We note that $\langle r_x^2\rangle^{1/2}_{I=0;1}=
\langle r_y^2\rangle^{1/2}_{I=0;1}$.
For comparison we also give the free 
space values of the root-mean-square radii.}
\end{table}

%%\end{document} %% preliminary end
%%%%%%%%%%%%%%%%%%%%%%%%%%%%%%%%%%%%%%%%%%%%%%%%%%%%%%%%%%%%%%%%%%%%%%%%%%%%%%%
\newpage
\vskip 3cm

\newpage

\newpage

%%%%%%%%%%%%%%%%%%%%%%%%%%%%%%%%%%%%%%%%%%%%%%%%%%%%%%%%%%%%%%%

\newpage

%%%%%%%%%%%%%%%%%%%%%%%%%%%%%%%%%%%%%%%%%%%%%%%%%%%%%%%%%%%%%%%%%%%%%

\newpage

%%%%%%%%%%%%%%%%%%%%%%%%%%%%%%%%%%%%%%%%%%%%%%%%%%%%%%%
\newpage

%%%%%%%%%%%%%%%%%%%%%%%%%%%%%%%%%%%%%%%%%%%%%%%%%%%%%%%%%%%%%%%%

\newpage

%%%%%%%%%%%%%%%%%%%%%%%%%%%%%%%%%%%%%%%%%%%%%%%%%%%%%%%%%%%%%%%%%
\newpage

%%%%%%%%%%%%%%%%%%%%%%%%%%%%%%%%%%%%%%%%%%%%%%%%%%%%%%%%%%%%%%%%%
\end{document}